\documentclass[aps,prl,reprint,longbibliography,superscriptaddress]{revtex4-2}
\usepackage{mathrsfs}
\usepackage{amsmath,gensymb}
\usepackage{amsfonts}
\usepackage{amssymb}
\usepackage{amsthm}
\usepackage{graphicx}
\usepackage{natbib}
\usepackage{color}
\usepackage{xcolor}
\usepackage{hyperref}
\usepackage{bm}
\usepackage[caption=false]{subfig}
\usepackage{verbatim}
\usepackage[normalem]{ulem}
\usepackage[version=4]{mhchem}

\begin{document}
\title{Proximity-induced orbital antiferromagnetism in Ising superconductors}

\author{G. A. Bobkov}
\email{gabobkov@mail.ru}
\affiliation{Moscow Institute of Physics and Technology, Dolgoprudny, 141700 Moscow region, Russia}

\author{V.A. Bobkov}
\affiliation{Moscow Institute of Physics and Technology, Dolgoprudny, 141700 Moscow region, Russia}

\author{T. Karabassov}
\affiliation{Moscow Institute of Physics and Technology, Dolgoprudny, 141700 Moscow region, Russia}
\affiliation{HSE University, 101000 Moscow, Russia}

\author{I.V. Bobkova}
\email{ivbobkova@mail.ru}
\affiliation{Moscow Institute of Physics and Technology, Dolgoprudny, 141700 Moscow region, Russia}

\author{A. A. Golubov}
\affiliation{Moscow Institute of Physics and Technology, Dolgoprudny, 141700 Moscow region, Russia}
\affiliation{HSE University, 101000 Moscow, Russia}

\begin{abstract} 
We predict a fundamentally new superconducting state in superconductor/antiferromagnet heterostructures with Ising spin--orbit coupling: proximity-induced orbital antiferromagnetism. In this state, the order parameter acquires a periodic phase modulation locked to the magnetic lattice, generating atomic-scale loop currents with opposite orbital moments on neighboring unit cells. Its emergence requires at least three nonequivalent magnetic sublattices per unit cell and finite spin--orbit coupling. Using NbSe$_2$/MnPS$_3$ as a concrete example, we combine first-principles and Bogoliubov--de Gennes calculations to demonstrate that the proximity-induced exchange field leads to robust phase modulation. Unlike FFLO and helical states, the phase gradient is atomic-scale, the state is current-carrying, and it remains uniquely stable over the full parameter range. The state manifests as characteristic finite-energy dips in the local density of states, accessible by STM.

\end{abstract}

\maketitle

{\it Introduction.}---The coexistence of superconductivity and magnetism yields a wealth of exotic superconducting states. The most prominent example is the partial conversion of singlet Cooper pairs into odd-frequency triplets in superconductor/ferromagnet (S/F) heterostructures, driven by the macroscopic exchange field \cite{Buzdin2005,Bergeret2005}. Such triplets have unveiled intriguing physics \cite{Buzdin2005,Bergeret2005,Bergeret2018_review} and provided a major impetus for superconducting spintronics \cite{Eschrig2015,Linder2015}.

Odd-frequency triplets also emerge in heterostructures with fully compensated antiferromagnets (zero net exchange field), alternating sign from site to site in mimicry of N\'eel order \cite{Bobkov2022,Bobkova2024_neel_review}. These N\'eel triplet Cooper pairs possess
a number of nontrivial properties \cite{Bobkov2023_anisotropy,Bobkov2023_impurities,Bobkov2023_oscillations,Chourasia2023,Bobkov2025,Bobkov2024_impurity} and enabling the spin-valve effect in S/AF/S heterostructures \cite{Johnsen_Kamra_2023,Bobkov2024_spinvalve}.

\begin{figure*}[!tbh]
\includegraphics[width=0.95\textwidth]{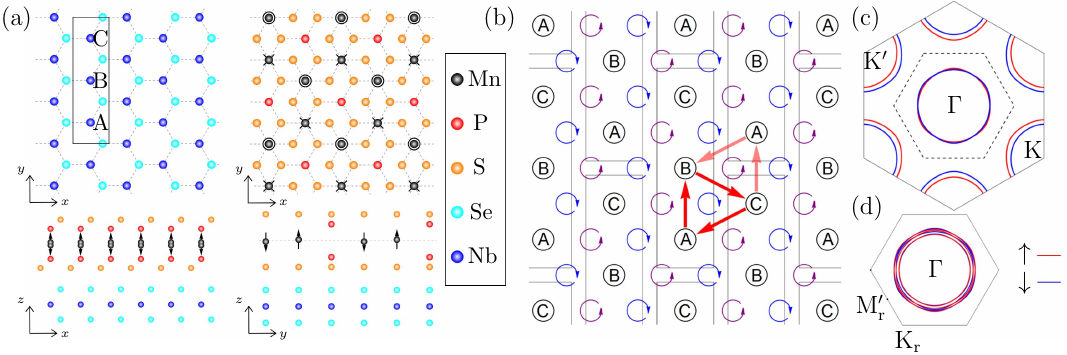}
\caption{(a) Top view of the ${\rm NbSe_2}$ and ${\rm MnPS_3}$ monolayers and side views of the ${\rm NbSe_2/MnPS_3}$ heterostructure in the $(x,z)$ and $(y,z)$ planes. Black arrows in the side views and circles/crosses in the top view indicate Mn magnetic moments aligned in $+z/-z$ directions respectively.
(b) Schematic of atomic-scale current loops with opposite circulation in the ${\rm NbSe_2}$ monolayer, induced by the phase difference of the superconducting order parameter on the Nb sites A, B, and C. Red arrows indicate the current flow between neighboring sites. 
(c) Brillouin zone and the spin-split Fermi surfaces of the isolated ${\rm NbSe_2}$ monolayer. The dashed curve marks the boundary of the folded Brillouin zone of the ${\rm NbSe_2/MnPS_3}$ heterostructure, arising from the magnetic superlattice. 
(d) Folded Brillouin zone of the ${\rm NbSe_2/MnPS_3}$ heterostructure, showing the  Fermi surfaces formed by the Fermi-level-crossing bands derived from ${\rm NbSe_2}$.} 
 \label{fig:1}
\end{figure*}

Here we predict a fundamentally new superconducting
phase emerging in heterostructures composed of an
Ising superconductor and a compensated
antiferromagnet. In this state the superconducting
order parameter acquires an atomic-scale phase
modulation locked to the magnetic lattice.
The superconducting condensate spontaneously develops
orbital antiferromagnetic order characterized by
staggered equilibrium loop currents and alternating
orbital magnetic moments (Fig.~\ref{fig:1}b). We term this state orbital
antiferromagnetic superconductivity.

Unlike FFLO superconductivity  \cite{Fulde1964,Larkin1965,Mironov2012,Mironov2018}, which exists only within
a narrow parameter window near the suppression of
superconductivity, the orbital-antiferromagnetic state
predicted here emerges as the unique equilibrium state
for arbitrary exchange fields and temperatures within
the superconducting phase. It also fundamentally differs from
the helical state \cite{Edelstein1989,Barzykin2002,Samokhin2004,Kaur2005,Dimitrova2007,Houzet2015,Rabinovich2019,Meng2019}, arising from coexisting spin--orbit coupling (SOC) and exchange field because it is current-carrying and remains uniquely stable over the full parameter range studied, whereas the homogeneous state is always metastable in the regime of the helical state existence. Both FFLO and helical inhomogeneous superconducting states rely on a
macroscopic exchange field, while the predicted orbital antiferromagnetic superconductivity can be produced even by a fully compensated antiferromagnet. 

Furthermore, the predicted state is distinct from the orbital altermagnetic state \cite{Pan2025} and orbital antiferromagnetism studied in correlated systems (underdoped cuprates \cite{Simon2002,Chakravarty2001,Chakravarty2001_2,Kee2002,Mook2001,Belyavsky2005}, SrRuO$_3$ \cite{Schroeter2004}, CeB$_6$ \cite{Ohkawa1985}, URu$_{2-x}$Fe$_x$Si$_2$ \cite{Kung2016}), where orbital order is normal-state driven and competes with superconductivity. By contrast, our atomic-scale loop currents exist only in the superconducting state and are carried by Cooper pairs.

Remarkably, two magnetic sublattices are insufficient
to generate orbital-antiferromagnetic superconductivity.
We show that the effect requires simultaneously
finite spin–orbit coupling and at least three inequivalent magnetic sublattices. We provide  a specific example: an S/AF heterostructure of superconducting NbSe$_2$ monolayer \cite{Xi2016,delaBarrera2018,Khestanova2018,Wickramaratne2023} and antiferromagnetic MnPS$_3$ monolayer \cite{Olsen2024,Long2020,Sun2019,Kim2019,Strasdas2023} (Fig.~\ref{fig:1}a). Combining first-principles electronic structure with Bogoliubov--de Gennes (BdG) calculations, we estimate the phase inhomogeneity and loop-current amplitude, and propose an scanning tunneling
microscopy (STM) detection scheme via a characteristic density-of-states feature.

{\it Orbital antiferromagnetism in a monolayer Ising superconductor.---}We analytically demonstrate that an atomically periodic exchange field in an Ising superconductor induces a phase modulation of the order parameter (OP) with the same periodicity, requiring at least three magnetic sublattices and finite spin--orbit coupling (SOC). Consider a triangular-lattice superconductor with arbitrary normal-state dispersion $\xi_\sigma(\bm p)$ ($\sigma=\uparrow,\downarrow$). SOC lifts spin degeneracy: $\xi_\sigma(\bm p)\neq\xi_{-\sigma}(\bm p)$, while time-reversal imposes $\xi_\sigma(\bm p)=\xi_{-\sigma}(-\bm p)$. This model describes monolayer monolayer
transition metal dichalcogenide superconductors such as NbSe$_2$. We assume an exchange field $\bm h_A\neq\bm h_B\neq\bm h_C$ on the three inequivalent Nb sites (A,B,C) per unit cell [Fig.~\ref{fig:1}(a),(c)]; below we show this is realized in NbSe$_2$/MnPS$_3$.

The system is described by the tight-binding Hamiltonian
\begin{align}
&\hat H = - \sum\limits_{ij,\sigma} c^\dagger_{i,\sigma} t_{ij,\sigma} c_{j,\sigma} -
\sum\limits_{i,\sigma} \mu c^\dagger_{i,\sigma} c_{i,\sigma} \nonumber \\
&+ \sum\limits_{i,\alpha,\beta} c^\dagger_{i,\alpha}(\bm h_i\bm\sigma)_{\alpha,\beta} c_{i,\beta} +\sum\limits_i\left[\Delta_i c_{i,\uparrow} c_{i,\downarrow}+H.c.\right] ,
\label{eq:hamiltonian_gen}
\end{align}
where $c_{i,\sigma}$ annihilates an electron at site $i$ with spin $\sigma$ and $\mu$ is the on-site energy. The proximity-induced exchange field is $\bm h_i = \left\{\bm h_A, \bm h_B, \bm h_C \right\}$. The OP $\Delta_i = \left\{\Delta_A, \Delta_B, \Delta_C \right\}$ is determined self-consistently via $\Delta_i = \lambda \langle c_{i,\downarrow}^S c_{i,\uparrow}^S \rangle$, with pairing constant $\lambda$; for the isolated monolayer it is homogeneous and equals $\Delta$. First-principles calculations indicate that Ising SOC in NbSe$_2$ imposes a strong perpendicular magnetic anisotropy on the Mn moments. Accordingly, we take $\bm h_{A,B,C} = h_{A,B,C}\bm e_z$. We assume $|\bm h_{A,B,C}| \ll \Delta$, which permits an analytical treatment. The numerical results for the realistic ${\rm NbSe_2/MnPS_3}$ heterostructure are given below. Expanding the Green's function to leading order in $|\bm h_{A,B,C}|/\Delta$~\cite{suppl} yields the sublattice-resolved correction to the OP:
\begin{align}
    &\delta \Delta_A =\frac{\lambda}{2} \sum \limits_{\bm Q, \sigma, \omega_m} \int \frac{d^2p}{3V}\big[  I_\sigma(\bm p,\bm p+\bm Q) h_{\bm Q} \nonumber \\
    &+ F_\sigma^\Delta (\bm p,\bm p+\bm Q) \delta \Delta_{\bm Q} + F_\sigma^{\Delta^*}(\bm p,\bm p+\bm Q) (\delta \Delta^*)_{\bm Q}  \big],
    \label{eq:Delta_a_general}
\end{align}
with analogous expressions for $\delta \Delta_{B,C}$. Here $\omega_m = \pi T (2m+1)$ is the Matsubara frequency, the integration runs over the Brillouin zone of the isolated monolayer (volume $V$), and $\bm Q = \left\{ 0, \bm K_1, \bm K_2 \right\}$ are the reciprocal vectors that fold the original Brillouin zone due to the induced magnetic superstructure. We use the notation $h_{\bm Q} = h_A + h_B e^{i \bm Q (\bm r_B-\bm r_A)} + h_C e^{i \bm Q (\bm r_C-\bm r_A)}$, where $\bm r_{A,B,C}$ are radius-vectors of the $A$, $B$ and $C$ sites of the same unit cell, and likewise for $\delta \Delta_{\bm Q}$ and $(\delta \Delta^*)_{\bm Q}$.
\begin{align}
    I_\sigma(\bm p_1, \bm p_2)=\sigma \Delta \frac{ \xi_\sigma (\bm p_1)-\xi_\sigma (\bm p_2)+2i\omega_m }{D_\sigma(\bm p_1, \bm p_2)},
    \label{I}
\end{align}
\begin{align}
    F_\sigma^\Delta(\bm p_1, \bm p_2)=-\frac{(\xi_\sigma (\bm p_1)+i \omega_m)(\xi_\sigma (\bm p_2)-2i\omega_m)}{D_\sigma(\bm p_1, \bm p_2)},
    \label{F_Delta}
\end{align}
\begin{align}
    F_\sigma^{\Delta^*}(\bm p_1, \bm p_2)=\frac{\Delta^2}{D_\sigma(\bm p_1, \bm p_2)},
    \label{F_Delta_star}
\end{align}
\begin{align}
D_\sigma(\bm p_1, \bm p_2)=(\Delta^2+\xi_\sigma^2 (\bm p_1)+\omega_m^2)(\Delta^2+\xi_\sigma^2 (\bm p_2)+\omega_m^2).    
\label{D}
\end{align}
From Eqs.~(\ref{I})--(\ref{D}), $I_\sigma(\bm p_1,\bm p_2,\omega_m)=-I_\sigma(\bm p_2,\bm p_1,-\omega_m)$ and $F_\sigma^{\Delta(\Delta^*)}(\bm p_1,\bm p_2,\omega_m)=F_\sigma^{\Delta(\Delta^*)}(\bm p_2,\bm p_1,-\omega_m)$. Consequently, fewer than three sublattices yield $\delta\Delta_{A,B,C}=0$: for two sublattices, $\bm Q$ and $-\bm Q$ are equivalent, causing the first term in Eq.~(\ref{eq:Delta_a_general}) to vanish. Additionally, the linear OP correction vanishes without SOC, as time-reversal and the absence of SOC imply $I_\sigma(\bm p_1,\bm p_2)=I_{-\sigma}(\bm p_1,\bm p_2)$, canceling upon spin summation.

For NbSe$_2$ the folding vectors are $\bm Q = \left\{ 0, \bm K, -\bm K \right\}$, with $\bm K = \frac{4\pi}{3a} \bm e_y$ connecting $\Gamma$ and $K$ points [Fig.~\ref{fig:1}(c)], and  $\bm r_B = a \bm e_y$, $\bm r_C = 2 a \bm e_y$, where $a$ is the nearest-neighbor Nb distance. The OP correction $\delta \check \Delta = \left(
\delta \Delta_A ,
\delta \Delta_B ,
\delta \Delta_C 
\right)^T$ then takes the explicit form
\begin{align}
\delta \check  \Delta = \frac{i I_h}{I_\Delta}(h_B-h_C, h_C-h_A, h_A - h_B)^T ,
    \label{Delta_explicit}
\end{align}
\begin{align}
    I_h =- \frac{\sqrt 3\lambda}{2} \sum \limits_{\sigma, \omega_m} \int \frac{d^2p}{3V}I_\sigma(\bm p,\bm p+K) ,
    \label{I_h}
\end{align}
\begin{align}
    I_\Delta =1- \frac{\lambda}{2} \sum \limits_{\sigma, \omega_m} \int \frac{d^2p}{V}\frac{F_\sigma^\Delta (\bm p,\bm p+K)-F_\sigma^{\Delta^*}(\bm p,\bm p+K)}{D_\sigma(\bm p,\bm p+K)} .
    \label{I_Delta}
\end{align}
Thus, in NbSe$_2$ the superconducting OP acquires a purely phase inhomogeneity driven by differences between exchange fields on distinct sublattices. The absence of a real part in the inhomogeneous correction is a consequence of the triangular lattice symmetry of NbSe$_2$; in a more general setting, the correction may have both real and imaginary components.

{\it \ce{NbSe2}/\ce{MnPS3} heterostructure.---} 
The effective field $h_{A,B,C}$ is extracted from the  band structure of the NbSe$_2$/MnPS$_3$ heterostructure in the normal (non-superconducting) state. The electronic structure calculations for the 1H-NbSe$_2$/MnPS$_3$ heterostructure are performed within the framework of density functional theory (DFT). All computations were carried out using the OpenMX software package \cite{Ozaki2003},\cite{Ozaki2004} employing the Perdew-Burke-Ernzerhof (PBE) functional within the generalized gradient approximation (GGA) \cite{Perdew1996}. Numerical integrations used an energy cutoff of 2700~eV, a vacuum layer $>15$~\AA, and a $14\times14\times1$ $k$-point grid. Van der Waals interactions were treated with DFT-D3 \cite{Grimme2010}.

\begin{figure}[!tbh]
\includegraphics[width=0.9\columnwidth]{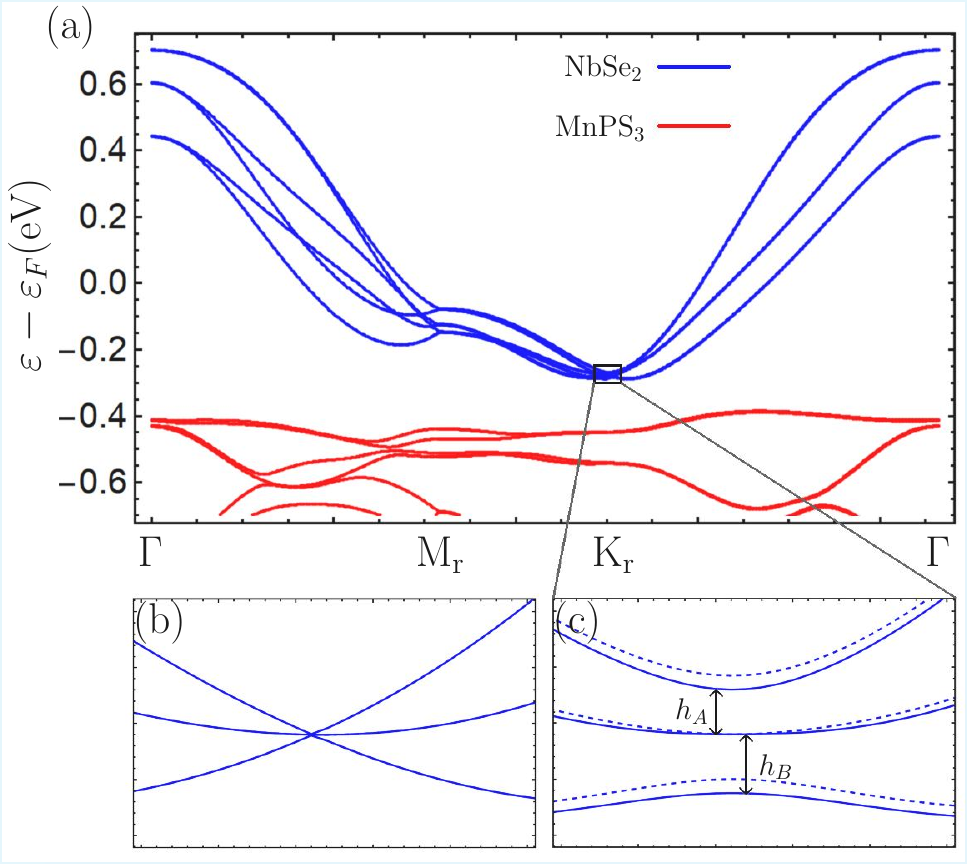}
\caption{(a) Electronic band structure of the NbSe$_2$/MnPS$_3$ heterostructure in the folded Brillouin zone [see Fig.~{fig:1}(d)]. Blue branches stem from NbSe$_2$; red branches are of MnPS$_3$ origin, which retains its insulating character in the heterostructure upon inclusion of interlayer charge transfer. (b)--(c) Zoom-in on the vicinity of the $K_r$ point (b) for a free-standing NbSe$_2$ monolayer and (c) for the NbSe$_2$/MnPS$_3$ heterostructure. Solid and dashed lines correspond to dpin-up and spin-down branches.} 
 \label{fig:2}
\end{figure}

The calculated band structure in the reduced Brillouin zone along the $\Gamma$--$M_r$--$K_r$--$\Gamma$ path [see Fig.~\ref{fig:1}(d)] is shown in Fig.~\ref{fig:2}. Since MnPS$_3$ is an insulator, only the conduction band originating from NbSe$_2$ is present at the Fermi level (shown in blue in Fig.~\ref{fig:2}). This band is nearly indistinguishable from that of an isolated NbSe$_2$ monolayer, as all bands originating from MnPS$_3$ lie far from the Fermi level (shown in red in Fig.~\ref{fig:2}). Therefore, we use the hopping parameters of isolated NbSe$_2$ monolayer as $t_{ij,\sigma}$, and describe the influence of the MnPS$_3$ solely through the induced exchange field $h_{A,B,C}$. We restrict ourselves to the nearest-neighbor hopping $t_{1,\sigma}=17.5e^{\pm 1.48\sigma i}$~meV, the next-nearest-neighbor hopping $t_{2,\sigma}=99.8$~meV, and the on-site energy $t_{0,\sigma}=31.4$~meV \cite{Aikebaier2022, Bobkov2024_vdWfirst}; further details on the parameter choice are provided in the Supplemental Material \cite{suppl}.  The complex $t_{1,\sigma}$ with opposite phases for A--B and A--C bonds captures Ising SOC. 

Unlike a ferromagnet, the antiferromagnet-induced exchange field does not cause a net conduction-band splitting; instead, it lifts degeneracy only at specific $k$-points. At $K_r$, symmetry forces all three branches to be degenerate without exchange \cite{suppl}, and the field lifts this degeneracy. In the folded BZ, the kinetic energy takes the form $E_{\sigma,\nu \nu'}(\bm p) = -\sum \limits_{\bm r} t_{ij,\sigma}e^{i \bm p \bm r}$, where the sites $i = (0,\nu)$ and $j=(\bm r, \nu')$ are labeled by the cell radius-vector $0$ and $\bm r$ and the sublattice index $\nu$ and $\nu'$, respectively. The triangular lattice symmetry allows the equivalent hopping terms to be grouped into triples $\left\{  t_{ij_1,\sigma}, t_{ij_2,\sigma}, t_{ij_3,\sigma} \right\}$  obtained by 120-degree rotations of the system. All these hopping elements are equal. It can be shown \cite{suppl} that if $\bm i$ and $\bm j$ belong to the same sublattice, then at the momentum $\bm p = \bm K_r$ corresponding to $K_r$-point $E_{\sigma,\nu \nu}(\bm K_r) = e^{2 i \pi k_1 /3} \sum \limits_{\bm r} t_{ij,\sigma}$ ($k_1$ is an integer). If $i$ and $j_n$ ($n\in \{1,2,3\}$) belong to different sublattices $\nu \neq \nu'$, then $E_{\sigma,\nu \nu'}(\bm K_r) = 0$ \cite{suppl}. Thus, for this particular momentum point $K_r$ all off-diagonal components of the sublattice matrix $E_{\sigma,\nu \nu'}(\bm p)$ vanish. 

Consequently, the electron eigenfunctions become strictly localized on a single sublattice. As a result, $h_{A,B,C}$ splits only the corresponding branch [Fig.~\ref{fig:2}(c)]. From DFT near $K_r$, we obtain $h_A=8$~meV, $h_B=-10.5$~meV, $h_C=0$. The asymmetry $|h_A|\neq|h_B|$ arises because Mn moments are asymmetrically positioned relative to the MnPS$_3$ layer center [Fig.~\ref{fig:1}(a)]. DFT gives perfect transparency ($D=1$), while in a real experiment achieving $D=1$ is practically impossible. For this reason in the following calculations, we will use the exchange field in the form $h_{A,B,C}=D \{+8,-10.5,0\}$meV, where $D\leq 1$ is the interface transparency. 

\begin{figure}[!tbh]
\includegraphics[width=0.9\columnwidth]{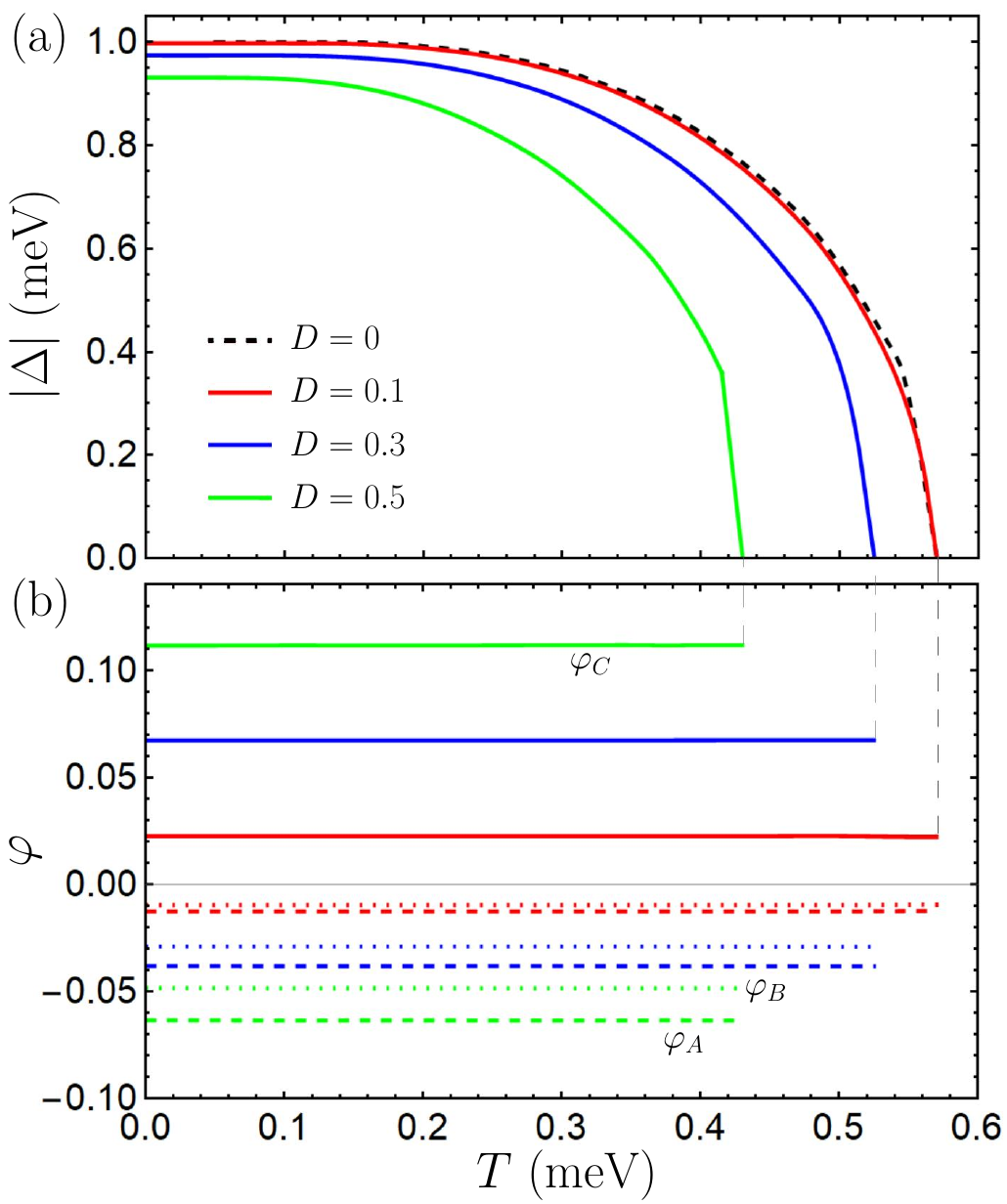}
\caption{Superconducting OP in the NbSe$_2$ monolayer with the effective exchange field $h_{A,B,C}=D\{+8,-10.5,0\}\,\text{meV}$ as a function of temperature. (a) OP amplitude (identical on all three magnetic sublattices); (b) OP phase on the three magnetic sublattices: $\varphi_A$ (dashed), $\varphi_B$ (dotted), $\varphi_C$ (solid).} 
 \label{fig:3}
\end{figure}

The OP is computed numerically via the BdG method \cite{suppl}. Figure~\ref{fig:3}(a) shows standard suppression of the OP amplitude with increasing $D$ (effective exchange field), stemming from singlet-to-triplet conversion \cite{Sarma1963,Bobkov2022} (conventional triplets for ferromagnets, N\'eel triplets for antiferromagnets). Here the effective field is ferrimagnetic, inducing a mixture of both; detailed triplet analysis is not directly relevant to the phase-inhomogeneous superconductivity problem and is deferred to the Supplemental Material \cite{suppl}.

Fig.~\ref{fig:3}(b) demonstrates a phase modulation $\varphi_{A,B,C}$ with magnetic periodicity for any nonzero exchange field. The phase is linear, to good accuracy, in $h_{A,B,C}$ (even beyond the $h\ll\Delta$ analytical limit) and temperature-independent, contrasting sharply with the FFLO state.

The obtained phase inhomogeneity is strong: the phase gradient $q_{\varphi}\sim0.1/a\sim10^9$~m$^{-1}$ (phase difference $\sim0.1$ between neighbors) far exceeds $q_c\sim\xi^{-1}\sim\Delta/(\hbar v_F)\sim10^4$~m$^{-1}$ associated with the critical current. Nevertheless, self-consistent BdG \cite{suppl} yields much smaller loop currents, e.g., $0.15I_c$ for $D=0.5$, $T=0.2$~meV ($I_c$ is NbSe$_2$ critical current), because quasiparticle wavefunctions, and hence the current between sites $i$ and $j$, average out atomic-scale phase variations over $\sim\xi$.

\begin{figure}[!tbh]
\includegraphics[width=0.9\columnwidth]{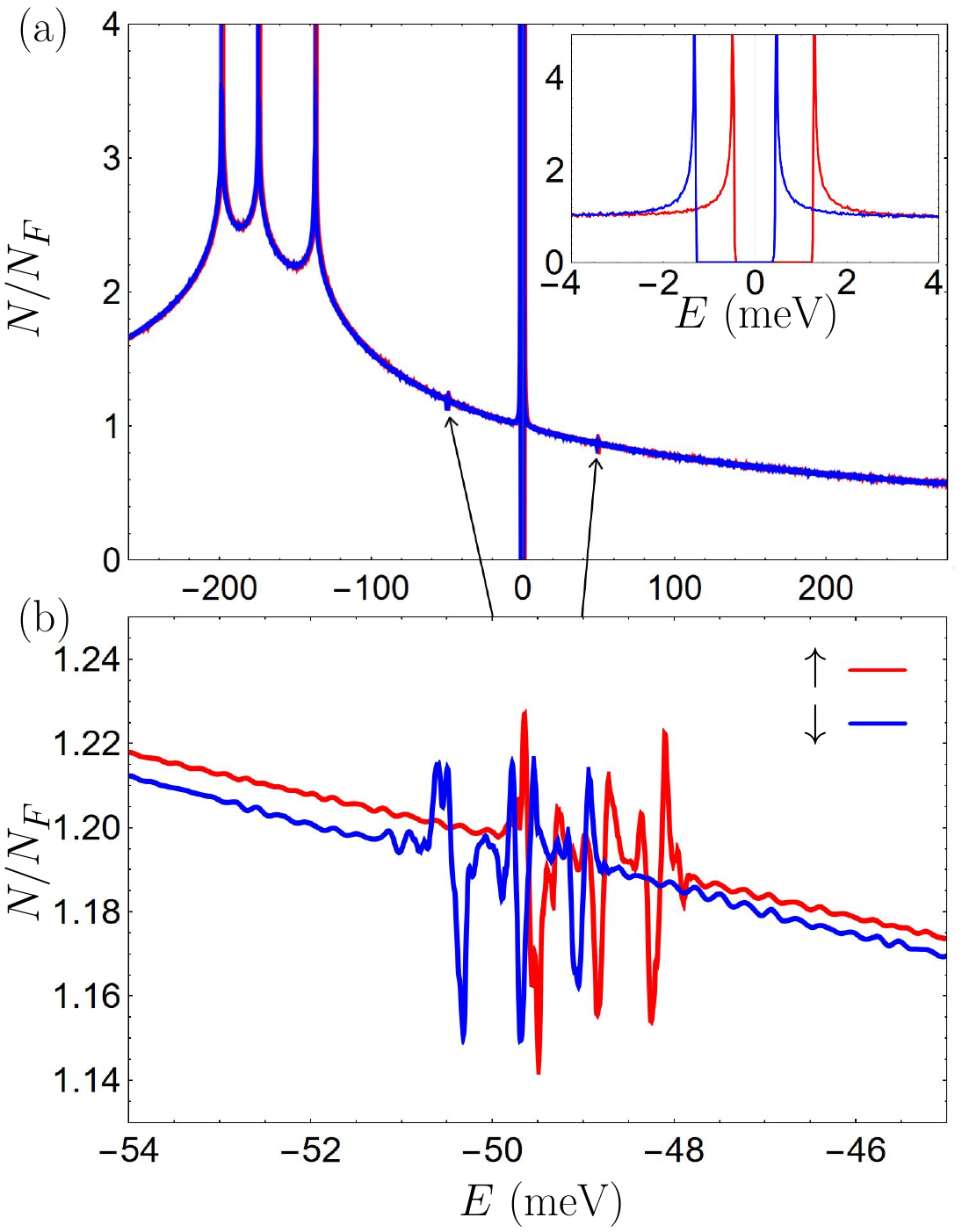}
\caption{(a) Spin-resolved LDOS (red: up, blue: down) of the superconducting monolayer in an effective exchange field. Inset: zoom of the superconducting gap at the Fermi-level. Finite-energy dips due to the periodic inhomogeneity of the superconducting OP are marked by arrows. (b) Zoom of one such dip region. $N_F$ is the normal-state LDOS at $E=0$.} 
 \label{fig:4}
\end{figure}

{\it Electronic LDOS.---}Fig.~\ref{fig:4} shows the calculated LDOS of the superconducting monolayer in the effective exchange field. Apart from finite-energy dips (marked by arrows), the LDOS is practically indistinguishable from that of isolated NbSe$_2$ monolayer. These dips arise from the periodic OP inhomogeneity and serve as an experimental signature. To explain their origin, consider a two-band model describing the vicinity of a band crossing of any two bands in the NbSe$_2$/MnPS$_3$ heterostructure:
\begin{align}
\hat H_m = \int \frac{d^2 p}{V} \left[ \sum \limits_{\nu} \varepsilon_{\nu}(\bm p)c_{\bm p, \sigma}^{\nu \dagger} c_{\bm p, \sigma}^{\nu} + t \sum \limits_{\nu \neq \nu'} c_{\bm p, \sigma}^{\nu \dagger} c_{\bm p, \sigma}^{\nu'} \right. \nonumber \\
\left. + \sum \limits_{\nu}(\Delta_{\nu}c_{\bm p, \uparrow}^{\nu \dagger}c_{\bm p, \downarrow}^{\nu \dagger} +H.c.) \right] ,  
    \label{ham_toy_model}
\end{align}
where $c_{\bm p,\sigma}^\nu$ is an electron annihilation operator in the band $\nu=1,2$, $\varepsilon_1(\bm p)=\xi(\bm p)$ and $\varepsilon_2(\bm p)=\xi(\bm p)+\Delta\xi$ are the diagonal in the sublattice space parts of the electron dispersion, and $t$ describes interaction between the bands accounting for the fact that electron is not localized at a given sublattice. Each of the bands has its own superconducting OP $\Delta_\nu$.

\begin{figure}[!tbh]
\includegraphics[width=0.9\columnwidth]{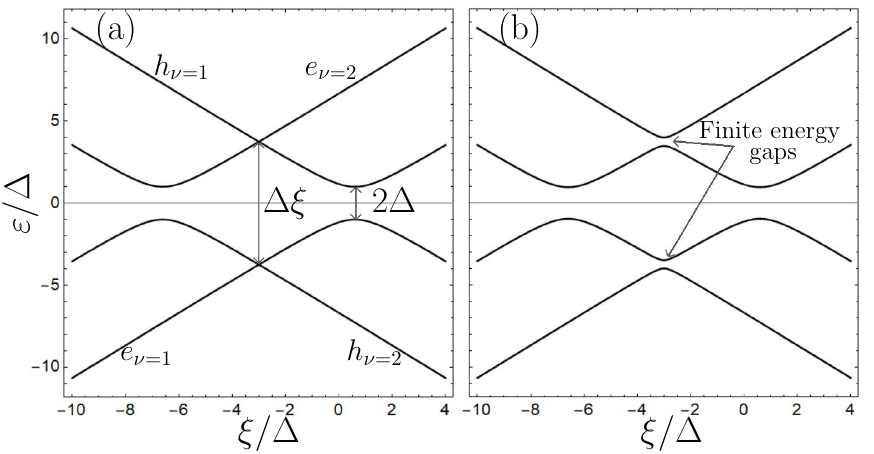}
\caption{Electronic spectra of the two-band model described by Eq.~(\ref{ham_toy_model}). The electron and hole branches of bands $\nu=1,2$ are labeled $e_{\nu=1,2}$ and $h_{\nu = 1,2}$, respectively. Parameters: $t=2\Delta$, $\Delta_1=\Delta=\Delta_2 e^{i\varphi}$, $\Delta\xi=6\Delta$. (a) $\varphi=0$, (b) $\varphi=1$.} 
 \label{fig:5}
\end{figure}

Diagonalizing this Hamiltonian yields the electronic spectrum $\varepsilon^{I(II)}$ of the superconducting monolayer in the vicinity of the  crossing point. Let's consider the crossing point  of the electronic branch $\varepsilon_1(\bm p)$ and the hole branch $-\varepsilon_2(\bm p)$, see Fig.~\ref{fig:5}. It occurs at  $\varepsilon \approx -\Delta\xi/2$ in the absence of the interaction between bands, $t=0$. Expanding to the first order of $\Delta_{1,2}/\Delta\xi$ and $t/\Delta\xi$ yields:
\begin{align}
    \varepsilon^{I,II}=-\Delta\xi/2\pm \frac{t|\Delta_1-\Delta_2|}{\Delta\xi}
\end{align}
Thus, a gap $2|\Delta_1-\Delta_2|t/\Delta\xi$ opens at finite energy $\approx\pm\Delta\xi/2$ when $\Delta_1\neq\Delta_2$, i.e., when the OP is periodically inhomogeneous. The gap energy is set by the normal-state band separation $\Delta\xi$, which is typically large compared to superconducting scales.

The three dips per spin in Fig.~\ref{fig:4} correspond to gaps at intersections of the three spectral branches of the three-atom-per-cell superconductor. Full gaps of the toy model become dips in the full calculation because branch crossings occur at slightly different energies for different momenta, smearing the gap.

Finite-energy dips are a characteristic signature of a periodically inhomogeneous OP, but they cannot distinguish phase from amplitude modulations. Unambiguous detection of phase inhomogeneity requires other probes (e.g., electromagnetic response), which is beyond this work's scope.

{\it Summary.---}We have established orbital-antiferromagnetic
superconductivity in Ising superconductor/antiferromagnet
heterostructures. In this state, the superconducting
condensate develops staggered equilibrium loop currents
and alternating orbital magnetic moments. The effect requires spin–orbit coupling and at least
three inequivalent magnetic sublattices. 
For the specific case of the NbSe$_2$/MnPS$_3$
heterostructure, first-principles calculations combined
with self-consistent Bogoliubov–de Gennes theory predict
phase differences of order
$0.1$ rad between neighboring sites  and loop currents reaching $\sim0.15I_c$.
The resulting state produces characteristic finite-energy
features in the local density of states, accessible by
scanning tunneling microscopy. Unlike the FFLO or helical states, the phase modulation persists as the unique superconducting state across all temperatures and exchange fields studied. Our findings establish a new paradigm for inhomogeneous superconductivity without a macroscopic exchange field.

\begin{acknowledgments}
G.A.B. and I.V.B. acknowledge the support from Theoretical Physics and Mathematics Advancement Foundation “BASIS” via the project  No. 23-1-1-51-1. The analytical and numerical calculations in the framework of the Green's function and BdG approaches were supported by the Russian Science Foundation via the project No. 24-12-00152. 
\end{acknowledgments}

\begin{widetext}

\section*{Supplemental Material for the Letter \\
	 ``Proximity-induced orbital antiferromagnetism in Ising superconductors''}

\subsection{Perturbative correction to the order parameter: Green's function approach}

Here we present the details of the derivation of the inhomogeneous correction to the superconducting order parameter to leading (first) order in the exchange field. Introducing the Nambu spinor $\check c_{\bm i} = (c_{{\bm i}\uparrow}, c_{\bm i\downarrow}, c_{\bm i\downarrow}^\dagger, -c_{\bm i\uparrow}^\dagger)^T$, we define the $4\times 4$ matrix Green's function in the direct product of spin and particle-hole spaces as
\begin{eqnarray}
\check G_{\bm i \bm j}(\tau_1-\tau_2) = - \tau_z \langle T_\tau \check c_{\bm i}(\tau_1) \check c_{\bm j}^\dagger(\tau_2) \rangle,
\label{eq:Green_Gorkov}
\end{eqnarray}
where $\langle T_\tau ... \rangle$ denotes imaginary time-ordered thermal averaging. Pauli matrices $\sigma_k$ and $\tau_k$ ($k=0,x,y,z$) act in spin and particle-hole spaces, respectively. The Green's function Eq.~(\ref{eq:Green_Gorkov}) satisfies
\begin{eqnarray}
\frac{d \check{G}_{\bm i \bm j}}{d \tau_1} = - \delta(\tau_1 -\tau_2) \delta_{\bm i \bm j} - \tau_z \langle T_{\tau } \frac{d \check{c_{\bm i}} (\tau_1)}{d \tau_1} \check{c^{\dag}_{\bm j}} (\tau_2) \rangle.
\label{eq:Green_derivative}
\end{eqnarray}
Using $d \check c_{\bm i}(\tau_1)/d \tau_1 = [\hat H , \check c_{\bm i}]$ with the Hamiltonian Eq.~(1) of the main text and expanding $\check G_{\bm i \bm j}(\tau_1 - \tau_2) = T \sum_{\omega_m} e^{-i \omega_m (\tau_1 - \tau_2)} \check G_{\bm i \bm j}(\omega_m)$ in fermionic Matsubara frequencies $\omega_m = \pi T(2m+1)$, we obtain the Gor'kov equation
\begin{align}
    (\hat H^0_{\bm i}-\delta \hat H_{\bm i})\check G_{\bm i\bm j}=\delta_{\bm i\bm j},
    \label{Gor'kov}
\end{align}
where
\begin{align}
    \hat H^0=i \omega_m \tau_z -\check \xi+\check \Delta , ~~~~ \check \xi \check{G}_{\bm i \bm j} = -\sum\limits_{\bm k} \hat t_{\bm i \bm k}\check{G}_{\bm k\bm j}-\mu \check{G}_{\bm i\bm j}.
\end{align}
Here $\hat t_{\bm i \bm k} = t_{ik,\uparrow}(1+\sigma_z)/2 + t_{ik,\downarrow}(1-\sigma_z)/2$ and $\check \Delta = \Delta i \tau_y$ describe the isolated superconducting monolayer, while
\begin{align}
    \delta \hat H_{\bm i}=\tau_z \bm \sigma \bm h_{\bm i}- \delta \check \Delta_{\bm i}
\end{align}
is the correction from the proximity-induced exchange field $\bm h_i = \left\{\bm h_A, \bm h_B, \bm h_C \right\}$ and $\delta \check \Delta_{\bm i} = \delta \Delta_{\bm i} \tau_+ + (\delta \Delta_{\bm i})^*\tau_-$ with $\tau_\pm = (\tau_x \pm i \tau_y)/2$. To first order in $h_{A,B,C}$ and $\delta \Delta \propto h$, the solution of Eq.~(\ref{Gor'kov}) reads
\begin{align}
    \check G_{\bm i\bm j}=\check G^0_{\bm i\bm j}+\sum_k G_{\bm i\bm k}^0 \delta \hat H_{\bm k} G_{\bm k\bm j}^0 ,
    \label{G_full_1}
\end{align}
where $\check G^0_{\bm i\bm j}$, the Green's function of the isolated monolayer, obeys $\hat H^0_{\bm i} \check G_{\bm i\bm j}^0=\delta_{\bm i\bm j}$ and can be expressed as
\begin{align}
    \check G_{\bm i\bm j}^0=\int \frac{d^2p}{V} \check G^0(\bm p)e^{i\bm p (\bm i-\bm j)}.
\end{align}
The integration runs over the Brillouin zone of the isolated monolayer (volume $V$). $\check G^0 (\bm p)$ is diagonal in spin space; its spin-$\sigma$ component is
\begin{align}
    \check G^0_\sigma(\bm p)=\frac{-1}{\Delta^2+\xi_\sigma^2(\bm p)+\omega_m^2}\left(\begin{array}{cc}
\xi_\sigma(\bm p)+i\omega_m &  \Delta \\
-\Delta &  \xi_\sigma(\bm p)-i\omega_m 
\end{array}\right),
\end{align}
with the normal-state dispersion $\xi_\sigma(\bm p)$ determined by the hopping elements $t_{ij,\sigma}$. In momentum space, Eq.~(\ref{G_full_1}) becomes
\begin{align}
    \check G_{\bm i\bm j}=\check G^0_{\bm i\bm j}+\sum_{\bm k} \int \frac{d^2p_1d^2p_2}{V^2} 
    \check G^0(\bm p_1)\delta H_{\bm k} \check G^0(\bm p_2)e^{i\bm p_1(\bm i-\bm k)+i\bm p_2(\bm k-\bm j)}.
\end{align}
The three sublattices contribute additively. Focusing on sublattice A, we write
\begin{align}
    \check G_{\bm i\bm j}^A =\check G^0_{\bm i\bm j}+\sum_{\bm k \in A} \int \frac{d^2p_1d^2p_2}{V^2} \check  G^0(\bm p_1)[\tau_z\sigma_z h_A -\delta \check \Delta_A]\check  G^0(\bm p_2)e^{i\bm p_1(\bm i-\bm k)+i\bm p_2(\bm k-\bm j)}.
    \label{G_A_1}
\end{align}
Using the identity
\begin{align}
    \sum_{\bm k \in A}e^{i\bm p\bm k}=\frac{V^2}{3}\sum_{\{\bm Q\}} \delta(\bm p-\bm Q),
\end{align}
where $\bm Q = \left\{ 0, \bm K_1, \bm K_2 \right\}$ are the reciprocal vectors that fold the original Brillouin zone due to the magnetic superstructure, we obtain from Eq.~(\ref{G_A_1})
\begin{align}
    \check G_{\bm i\bm j}^A =\check G^0_{\bm i\bm j}+\sum_{\{\bm Q\}} \int \frac{d^2p}{3V} \check  G^0(\bm p)[\tau_z\sigma_z h_A -\delta \check \Delta_A]\check  G^0(\bm p+\bm Q)e^{i(\bm p \bm i-(\bm p+\bm Q)\bm j)}.
\end{align}
The perturbation of the local Green's function $\delta\check G_{\bm i\bm i}^A$ then reads
\begin{align}
    \delta\check G_{\bm i\bm i}^A=\sum_{\{\bm Q\}} \int \frac{d^2p}{3V} \check G^0(\bm p)[\tau_z\sigma_z h_A -\delta \check \Delta_A] \check  G^0(\bm p+\bm Q)e^{-i\bm Q\bm i}.
    \label{delta_G_local}
\end{align}
Evaluating the matrix product,
\begin{align}
    \check G^0_\sigma(\bm p_1)[\tau_z\sigma_z h_A - \delta \check \Delta_A]\check G^0_\sigma(\bm p_2)=\frac{h_A\sigma}{D_\sigma} \left(\begin{array}{cc}
G_{\sigma}^h &  F_{\sigma}^h \\
\tilde F_{\sigma}^h &  \tilde G_{\sigma}^h 
\end{array}\right) + \delta \Delta_A \left(\begin{array}{cc}
G_\sigma^\Delta &  F_\sigma^\Delta \\
\tilde F_\sigma^\Delta &  \tilde G_\sigma^\Delta 
\end{array}\right) + (\delta \Delta_A)^* \left(\begin{array}{cc}
G_\sigma^{\Delta^*} &  F_\sigma^{\Delta^*} \\
\tilde F_\sigma^{\Delta^*} &  \tilde G_\sigma^{\Delta^*} 
\end{array}\right) ,
\end{align}
where
\begin{align}
&D_\sigma(\bm p_1,\bm p_2)=(\Delta^2+\xi_\sigma^2 (\bm p_1)+\omega_m^2)(\Delta^2+\xi_\sigma^2 (\bm p_2) +\omega_m^2),    \nonumber \\ 
&G_{\sigma}^h(\bm p_1,\bm p_2) =\Delta^2+(\xi_\sigma (\bm p_1)+i\omega_m)(\xi_\sigma (\bm p_2)+i\omega_m), \nonumber \\
&\tilde G_{\sigma}^h(\bm p_1,\bm p_2) =-\Delta^2-(\xi_\sigma (\bm p_1)-i\omega_m)(\xi_\sigma (\bm p_2)-i\omega_m),  \nonumber \\
&F_{\sigma}^h(\bm p_1,\bm p_2)=\Delta (\xi_\sigma (\bm p_1)-\xi_\sigma (\bm p_2)+2i\omega_m), \nonumber \\
&\tilde F_{\sigma}^h(\bm p_1,\bm p_2)=\Delta (\xi_\sigma (\bm p_1)-\xi_\sigma (\bm p_2)-2i\omega_m); \nonumber \\
&G_{\sigma}^\Delta(\bm p_1,\bm p_2) =\frac{\Delta(\xi_\sigma (\bm p_1)+i\omega_m)}{D_\sigma(\bm p_1,\bm p_2)}, \nonumber \\
&\tilde G_{\sigma}^\Delta(\bm p_1,\bm p_2) =\frac{\Delta (\xi_\sigma (\bm p_2)-i\omega_m)}{D_\sigma(\bm p_1,\bm p_2)},  \nonumber \\
&F_{\sigma}^\Delta(\bm p_1,\bm p_2)=-\frac{(\xi_\sigma (\bm p_1)+i\omega_m)(\xi_\sigma (\bm p_2)-i\omega_m)}{D_\sigma(\bm p_1,\bm p_2)}, \nonumber \\
&\tilde F_{\sigma}^\Delta(\bm p_1,\bm p_2)=-\frac{\Delta^2}{D_\sigma(\bm p_1,\bm p_2)}; \nonumber \\
&G_{\sigma}^{\Delta^*}(\bm p_1,\bm p_2) =\frac{\Delta(\xi_\sigma (\bm p_2)+i\omega_m)}{D_\sigma(\bm p_1,\bm p_2)}, \nonumber \\
&\tilde G_{\sigma}^{\Delta^*}(\bm p_1,\bm p_2) =\frac{\Delta (\xi_\sigma (\bm p_1)-i\omega_m)}{D_\sigma(\bm p_1,\bm p_2)},  \nonumber \\
&F_{\sigma}^{\Delta^*}(\bm p_1,\bm p_2)=\frac{\Delta^2}{D_\sigma(\bm p_1,\bm p_2)}, \nonumber \\
&\tilde F_{\sigma}^{\Delta^*}(\bm p_1,\bm p_2)=\frac{(\xi_\sigma (\bm p_1)-i\omega_m)(\xi_\sigma (\bm p_2)+i\omega_m)}{D_\sigma(\bm p_1,\bm p_2)} .
\label{def}
\end{align}

Finally, from Eq.~(\ref{delta_G_local}) the perturbation of the local anomalous Green's function at sublattice $A$ takes the form
\begin{align}
    \delta F_\sigma^A=\sum_{\{\bm Q\}}\int \frac{d^2p}{3 V}\big[  I_\sigma(\bm p,\bm p+\bm Q)h_{\bm Q}+ 
    F_\sigma^\Delta(\bm p,\bm p+\bm Q)\delta \Delta_{\bm Q}+ 
    F_\sigma^{\Delta^*}(\bm p,\bm p+\bm Q)(\delta \Delta^*)_{\bm Q}  \big]
    \label{delta_F_A}
\end{align}
with
\begin{align}
    I_\sigma(\bm p_1, \bm p_2)=\sigma\frac{ F_\sigma^h(\bm p_1, \bm p_2)}{D_\sigma(\bm p_1, \bm p_2)}.
    \label{def_I}
\end{align}
Here $h_{\bm Q} = h_A + h_B e^{i \bm Q (\bm r_B-\bm r_A)} + h_C e^{i \bm Q (\bm r_C-\bm r_A)}$, $\delta \Delta_{\bm Q} = \delta \Delta_A + \delta \Delta_B e^{i \bm Q (\bm r_B-\bm r_A)} + \delta \Delta_C e^{i \bm Q (\bm r_C-\bm r_A)}$ and  $(\delta \Delta)^*_{\bm Q} = (\delta \Delta_A)^* + (\delta \Delta_B)^* e^{i \bm Q (\bm r_B-\bm r_A)} + (\delta \Delta_C)^* e^{i \bm Q (\bm r_B-\bm r_C)}$, where $\bm r_{A,B,C}$ are radius-vectors of the $A$, $B$ and $C$ sites of the same unit cell.

Substituting Eq.~(\ref{delta_F_A}) into the self-consistency condition $\delta \Delta_A = \lambda \langle c_{A,\downarrow}^S c_{A,\uparrow}^S \rangle = (\lambda/2)\sum \limits_{\omega_m,\sigma}\delta F_\sigma^A (\omega_m)$ yields
\begin{align}
    \delta \Delta_A =\frac{\lambda}{2} \sum \limits_{\bm Q, \sigma, \omega_m} \int \frac{d^2p}{3V}\big[  I_\sigma(\bm p,\bm p+\bm Q) h_{\bm Q} 
    + F_\sigma^\Delta (\bm p,\bm p+\bm Q) \delta \Delta_{\bm Q} + F_\sigma^{\Delta^*}(\bm p,\bm p+\bm Q) (\delta \Delta^*)_{\bm Q}  \big],
    \label{eq:Delta_a_general_suppl}
\end{align}
which is Eq.~(2) of the main text.

Time-reversal symmetry of the normal-state spectrum of the isolated monolayer, $\xi_\sigma(\bm p) = \xi_{-\sigma}(-\bm p)$, together with Eqs.~(\ref{def}) and (\ref{def_I}) implies
\begin{align}
    I_\sigma(\bm p_1, \bm p_2, \omega_m)&=-I_{-\sigma}(-\bm p_1, -\bm p_2, \omega_m),  \nonumber \\
    I_\sigma(\bm p_1, \bm p_2, \omega_m)&=-I_{\sigma}(\bm p_2, \bm p_1, -\omega_m), \nonumber \\
    F_\sigma^\Delta (\bm p_1, \bm p_2, \omega_m) &= F_\sigma^\Delta (\bm p_2, \bm p_1, -\omega_m), \nonumber \\
     F_\sigma^{\Delta^*} (\bm p_1, \bm p_2, \omega_m) &= F_\sigma^{\Delta^*} (\bm p_2, \bm p_1, -\omega_m) .
\end{align}

Specializing the above relations to NbSe$_2$, where the folding vectors are $\bm Q = \left\{ 0, \bm K, -\bm K \right\}$ with $\bm K = \frac{4\pi}{3a}$, and $\bm r_B = a \bm e_y$, $\bm r_C = 2 a \bm e_y$, Eq.~(\ref{eq:Delta_a_general_suppl}) can be written explicitly in terms of $\delta \check \Delta = \left(
\delta \Delta_A ,
\delta \Delta_B ,
\delta \Delta_C 
\right)^T$ as
\begin{align}
    \delta\Delta_A &= \frac{\lambda}{2} \sum \limits_{\sigma, \omega_m} \int \frac{d^2p}{3V} \Bigl[-\sqrt 3 i I_\sigma(\bm p, \bm p+\bm K)(h_B -h_C)  + \bigl(F_\sigma^\Delta(\bm p, \bm p) + 2 F_\sigma^\Delta(\bm p, \bm p+ \bm K)\bigr) \delta \Delta_A \nonumber \\
    &+ \bigl(F_\sigma^\Delta(\bm p, \bm p) - F_\sigma^\Delta(\bm p, \bm p+ \bm K)\bigr)\bigl( \delta \Delta_B + \delta \Delta_C \bigr)+ \bigl(F_\sigma^{\Delta^*}(\bm p, \bm p) + 2 F_\sigma^{\Delta^*}(\bm p, \bm p+ \bm K)\bigr) \bigl(\delta \Delta_A \bigr)^* \nonumber \\
    &+ \bigl(F_\sigma^{\Delta^*}(\bm p, \bm p) - F_\sigma^{\Delta^*}(\bm p, \bm p+ \bm K)\bigr)\bigl( \delta \Delta_B + \delta \Delta_C \bigr)^* \Bigr].
    \label{Delta_explicit_suppl}
\end{align}
$\Delta_B$ and $\Delta_C$ satisfy the same equation under the cyclic substitutions $ABC \to BCA$ and $ABC \to CAB$, respectively.

The first term on the right-hand side of Eq.~(\ref{Delta_explicit_suppl})---the generator of the inhomogeneous correction to the OP---is purely imaginary. Hence $\delta \Delta_{A,B,C} = -(\delta \Delta_{A,B,C})^*$ are purely imaginary, i.e., the OP acquires a purely phase inhomogeneity. Using this fact and solving the coupled linear system Eq.~(\ref{Delta_explicit_suppl}) for $\left\{ \delta \Delta_{A}, \delta \Delta_{B}, \delta \Delta_{C} \right\}$ yields Eq.~(7) of the main text.

\subsection{Evaluation of the effective exchange field from DFT-calculated spectra of \ce{NbSe2}/\ce{MnPS3} heterostructure.} 

Unlike a ferromagnet, the exchange field $h_{A,B,C}$ induced by the antiferromagnet produces no exchange splitting of the conduction band and hence cannot be readily extracted from the DFT-calculated band structure. To obtain the exchange field we examine the vicinity of the $K_r$ point (see Figs.~3(b)--(c) of the main text). The symmetry of the \ce{NbSe2} crystal lattice dictates that at this point of the folded BZ all three branches of the band structure are degenerate in the absence of an exchange field [Fig.~3(b)]. To demonstrate this, we consider the first hopping term of Hamiltonian (1) of the main text, which corresponds to the kinetic energy $\hat H_{\rm kin}$, in the momentum representation $c_{i,\sigma} = \int (d^2p/V_r)\, c_{\bm p, \sigma}^\nu e^{i \bm p (\bm r_i+\bm r_\nu)}$, where the site index $i=(\bm r_i, \nu)$ comprises the cell radius-vector $\bm r_i$ and the sublattice index $\nu$. The basis vectors of the triangular lattice can be chosen as $\bm a = a(0,1,0)^T$ and $\bm b = a(\sqrt 3/2, 1/2, 0)^T$. The radius vectors of the three sites within a unit cell are then $\bm r_A = 0$, $\bm r_B = \bm a$, $\bm r_C = 2 \bm a$, and $V_r$ denotes the volume of the folded BZ:
\begin{align}
\hat H_{\rm kin} = \sum \limits_{\nu \nu'}\int \frac{d^2 p}{V_r} E_{\sigma, \nu \nu'}(\bm p)\, c_{\bm p, \sigma}^{\nu \dagger} c_{\bm p, \sigma}^{\nu '} . 
    \label{H_kin}
\end{align}
Here $E_{\sigma,\nu \nu'}(\bm p) = -\sum \limits_{\bm r_j - \bm r_i} t_{ij,\sigma}e^{i \bm p (\bm r_j - \bm r_i +\bm r_{\nu'}-\bm r_{\nu})}$, with $\bm r_j - \bm r_i = (\bm a + \bm b)k_1 + 3 \bm a k_2$ and $k_{1,2}$ integer. The reciprocal lattice basis vectors are $\tilde {\bm a} = (2\pi/a)(2/\sqrt 3, 0, 0)^T$ and $\tilde {\bm b} = (2\pi/a)(-1/\sqrt 3, 1, 0)^T$; the momentum corresponding to the $K_r$ point is $\bm K_r = \tilde {\bm a}/3$.

The triangular lattice symmetry allows the equivalent hopping terms to be grouped into triples \\ $\{ t_{ij_1,\sigma}, t_{ij_2,\sigma}, t_{ij_3,\sigma} \}$ related by $120^\circ$ rotations. All three elements in such a triple are equal. Evaluating at $\bm p = \bm K_r = \tilde {\bm a}/3$, we have $\bm K_r (\bm r_{j_2} - \bm r_i) = (\tilde {\bm b}/3) (\bm r_{j_1} - \bm r_i)$ and $\bm K_r (\bm r_{j_3} - \bm r_i) = -(\tilde {\bm a}/3 + \tilde {\bm b}/3) (\bm r_{j_1} - \bm r_i)$. Consequently, $e^{i\bm K_r (\bm r_{j_1} - \bm r_i)} = e^{i \bm K_r (\bm r_{j_2} - \bm r_i)} = e^{i \bm K_r (\bm r_{j_3} - \bm r_i)} = e^{2\pi i k_1/3}$. When sites $i$ and $j$ belong to the same sublattice, $E_{\sigma,\nu \nu}(\bm K_r) = e^{2 i \pi k_1 /3} \sum \limits_{\bm r_j - \bm r_i} t_{ij,\sigma} $. When $i$ and $j_n$ ($n\in\{1,2,3\}$) belong to different sublattices $\nu \neq \nu'$, for a given triple of identical hoppings one finds $e^{i\bm K_r (\bm r_\nu - \bm r_{\nu'})_1} = 1$, $e^{i \bm K_r (\bm r_\nu - \bm r_{\nu'})_2} = e^{i(\tilde {\bm b}/3) (\bm r_\nu - \bm r_{\nu'})_1} = e^{2i \pi/3}$, and $e^{i\bm K_r (\bm r_\nu - \bm r_{\nu'})_3} = e^{-i(\tilde {\bm b}/3 + \tilde {\bm a}/3) (\bm r_\nu - \bm r_{\nu'})_1} = e^{-2i \pi/3}$, which yields $E_{\sigma,\nu \nu'}(\bm K_r) = 0$. Thus all off-diagonal components of the sublattice matrix $E_{\sigma,\nu \nu'}(\bm p)$ vanish at $\bm K_r$.

Hence Hamiltonian (\ref{H_kin}) becomes diagonal in the sublattice space, implying that the electron eigenfunctions are strictly localized on individual sublattices. As a result, an exchange field $\bm h_{A,B,C}$ acting on a given sublattice splits the energy of only the corresponding branch [Fig.~2(c) of the main text].

\subsection{Exact numerical treatment via the Bogoliubov-de Gennes equations}

As our calculations show, the effective exchange field induced by MnPS$_3$ in NbSe$_2$ lies beyond the $|h| \ll \Delta$ regime. Consequently, for the parameters corresponding to the real heterostructure the superconducting order parameter, spontaneous currents, and density of states were obtained by solving the Bogoliubov--de Gennes (BdG) equations numerically. The details of the numerical procedure are described below. 

We diagonalize the Hamiltonian (1) of the main text by the Bogoliubov transformation:
\begin{align}
c_{\bm i,\sigma}=\sum\limits_n u_{n,\sigma}^{\bm i}\hat b_n+v^{\bm i *}_{n,\sigma}\hat b_n^\dagger . 
\label{bogolubov}
\end{align}
The resulting BdG equations take the form:
\begin{align}
 \sigma \Delta_{\bm i} v^{ \bm i}_{n,-\sigma} - \sum\limits_{ \bm i'} t_{\bm i \bm i',\sigma} u^{\bm i'}_{ n, \sigma} + h_i \sigma u^{\bm i}_{ n, \sigma} & = \varepsilon_n u_{n,\sigma}^{\bm i} \nonumber \\  
\sigma \Delta_{ \bm i}^* u^{\bm i}_{n,-\sigma} - \sum\limits_{ \bm i'} t_{\bm i \bm i',\sigma}^* v^{\bm i'}_{ n, \sigma} + h_i \sigma v^{\bm i}_{ n, \sigma} & = -\varepsilon_n v_{n,\sigma}^{\bm i}, 
\label{bdg}
\end{align}
Using the solutions of the BdG equations, the superconducting OP at site $\bm i$ and current flowing between sites $\bm i$ and $\bm i'$ are given by:
\begin{align}
    \Delta_{\bm i}=\lambda \sum_{n} \bigl( u_{n,\uparrow}^{ \bm i}v_{n,\downarrow}^{ \bm i*}f_n+u_{n,\downarrow}^{ \bm i}v_{n,\uparrow}^{ \bm i*}(1-f_n)\bigr)
\end{align}

\begin{align}
    \bm j_{\bm i \to\bm i'}=e\sum_{n,\sigma} i  \left[(t_{\bm i \bm i',\sigma}u_{n,\sigma}^{ \bm i} u_{n,\sigma}^{\bm i'*}-t^*_{\bm i \bm i',\sigma}u_{n,\sigma}^{ \bm i'} u_{n,\sigma}^{\bm i*})f_n +(t^*_{\bm i \bm i',\sigma}v_{n,\sigma}^{ \bm i*} v_{n\sigma}^{\bm i'}-t_{\bm i \bm i',\sigma}v_{n,\sigma}^{ \bm i'*} v_{n,\sigma}^{\bm i})(1-f_n)\right],
\end{align}
where $f_n=\frac{1}{e^{\varepsilon_n/T}+1}$ is a Fermi–Dirac distribution.
The local electronic density of states at site $i$ is obtained from the solutions of the BdG equations as
\begin{align}
    N(\varepsilon)=\sum_{n\sigma} |u_{n,\sigma}^{\bm i}|^2 \delta(\varepsilon-\varepsilon_n) .
\end{align}

To describe the band structure of NbSe$_2$, we restrict ourselves to two nearest-neighbor hopping terms, $t_{1,2}$, and the on-site energy $t_0$. This choice is motivated by the fact that calculating local current loops in a system with many long-range hopping elements seems physically meaningless. At the same time, a model with six nearest-neighbor hoppings was considered in previous studies \cite{Aikebaier2022, Bobkov2024_vdWfirst}. It is worth noting that the six-hopping model reproduces the actual DFT-calculated band structure of NbSe$_2$ only marginally better than the two-hopping model; the errors of both models are of the same order. In this sense, using only two hopping terms is fully justified and does not involve any crude approximations. It is also important to note that since the Fermi surface of NbSe$_2$ around the $\Gamma$ point transforms almost exactly into the Fermi surface around the $\rm K$ point when shifted by the $ K$ vector, the value of the integral $I_{\sigma}(\bm p,\bm p+K)$ is highly sensitive to the specific details of the model. Consequently, the resulting phase inhomogeneity may differ noticeably in quantitative terms between the two-hopping model, the six-hopping model, and a hypothetical model that perfectly describes the band structure of NbSe$_2$—although the very existence of the inhomogeneity and its overall structure will remain unchanged.

\subsection{Triplet correlations in \ce{NbSe2}/\ce{MnPS3} heterostructure.}

Here we present results for the amplitude and spatial structure of the {\it triplet} superconducting  correlations generated by the proximity-induced exchange field $\{h_A,h_B,h_C\}$ in the \ce{NbSe2} superconducting layer. The anomalous Green's function in Matsubara representation can be calculated as $F_{\bm i, \alpha \beta} = - \langle \hat c_{\bm i \alpha}(\tau) \hat c_{\bm i \beta}(0) \rangle$, where $\tau$ is the imaginary time. The component of this anomalous Green's function for a given Matsubara frequency $\omega_m = \pi T(2m+1)$ is calculated as follows: 
\begin{align}
F_{i,\alpha\beta}(\omega_m)= \sum\limits_n (\frac{  u_{n,\alpha}^{\bm i} v_{n,\beta}^{\bm i*}}{i \omega_m -\varepsilon_n}+\frac{ u_{n,\beta}^{\bm i} v_{n,\alpha}^{\bm i*}}{i \omega_m +\varepsilon_n})
\end{align}
Only off-diagonal in spin space components, corresponding to opposite-spin pairs, are nonzero for the case under consideration, when all magnetic moments are aligned along the same axis. The singlet (triplet) correlations are described by $F_{\bm i}^{s,t}(\omega_m) = F_{\bm i,\uparrow \downarrow}(\omega_m) \mp F_{\bm i,\downarrow \uparrow}(\omega_m)$. Please note that the  on-site triplet correlations are odd in Matsubara frequency, as it should be according to the general fermionic symmetry. Therefore we only consider the triplet correlations at the first positive Matsubara frequency $F_{i}^t(\omega_0) $. 

\begin{figure}[!tbh]
\includegraphics[width=0.6\columnwidth]{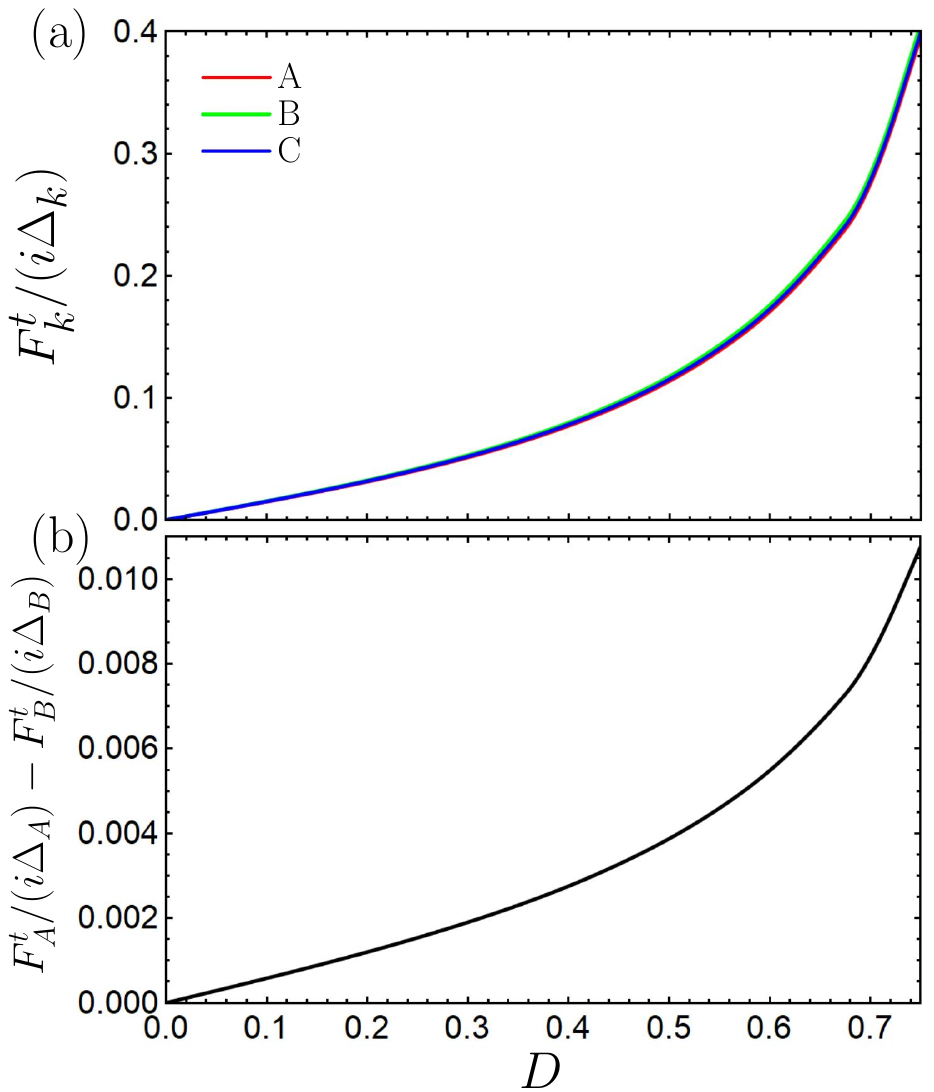}
\caption{(a) Sublattice-resolved amplitudes of triplet correlations in NbSe$_2$/MnPS$_3$ as functions of the interface transparency $D$. The triplet anomalous Green's functions are normalized by the complex order parameter on the corresponding sublattice to remove the additional spatial modulation imposed by the order parameter phase (see text). (b) Amplitude of the N\'eel triplet component $\tilde F_N^t = F_A^t/(i\Delta_A) - F_B^t/(i\Delta_B)$ versus $D$. Results are shown for the ferrimagnetic effective exchange field $h_{A,B,C}=D\{+8,-10.5,0\}$~meV.} 
 \label{fig:S1}
\end{figure}

\begin{figure}[!tbh]
\includegraphics[width=0.63\columnwidth]{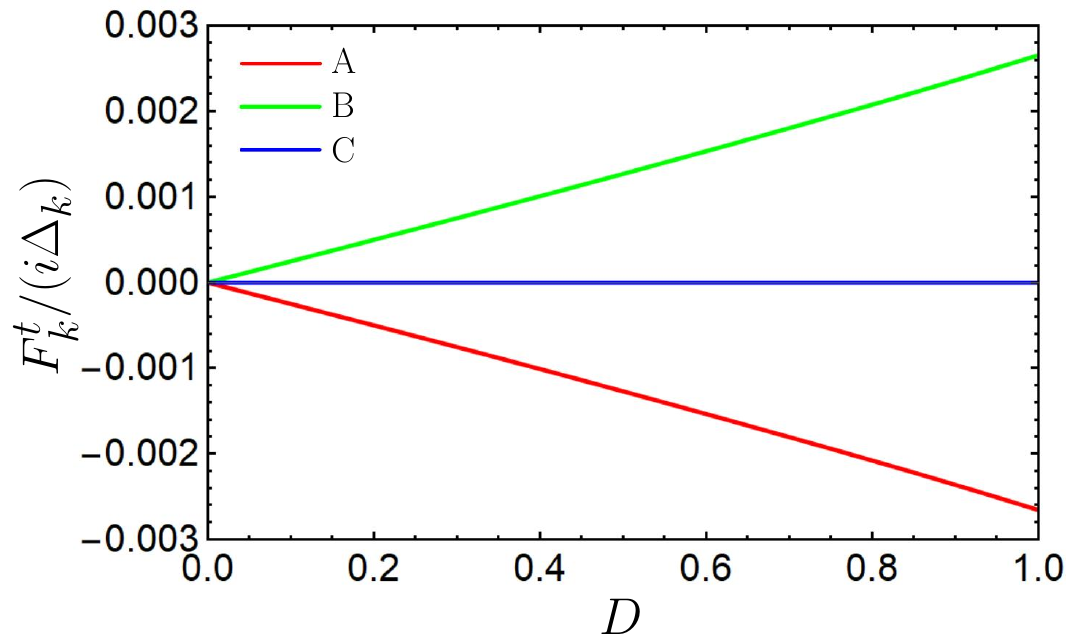}
\caption{Sublattice-resolved amplitudes of triplet correlations in NbSe$_2$/MnPS$_3$ as functions of the interface transparency $D$. Results are shown for the antiferromagnetic effective exchange field $h_{A,B,C}=D\{+8,-8,0\}$~meV.} 
 \label{fig:S2}
\end{figure}

\begin{figure}[!tbh]
\includegraphics[width=0.9\columnwidth]{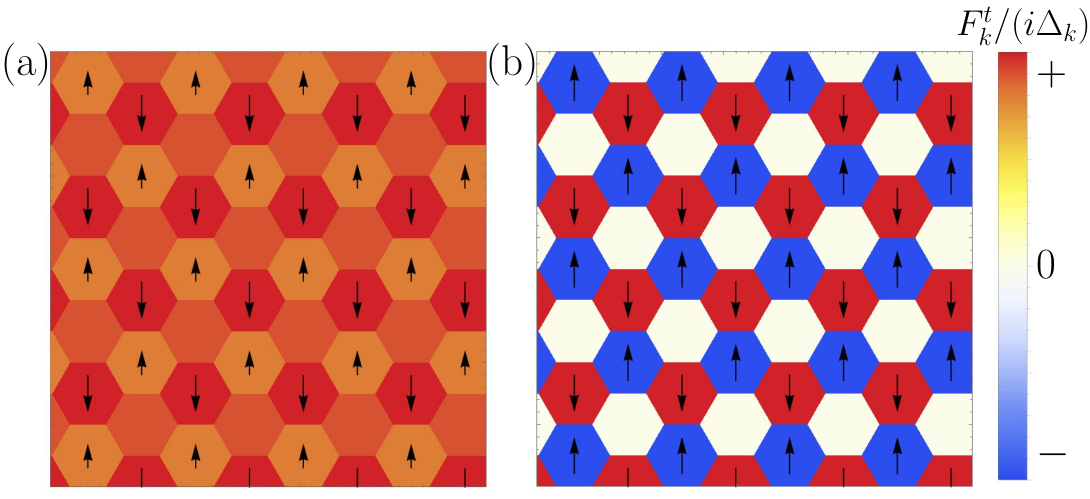}
\caption{Schematic (not to scale) of the spatial distribution of the triplet correlation ampli-
tude for (a) ferrimagnetic exchange field corresponding to Fig.~\ref{fig:S1} and (b) antiferromagnetic exchange field corresponding to Fig.~\ref{fig:S2}.} 
 \label{fig:S3}
\end{figure}

In Fig.~\ref{fig:S1}a, this quantity is plotted for all three sublattices as a function of the interface transparency $D$. The exchange field $h_{A,B,C}=D\{+8,-10.5,0\}$~meV is taken exactly as obtained from our DFT calculations for the NbSe$_2$/MnPS$_3$ heterostructure. It is well known that in the simplest cases of a homogeneous order parameter, the phase of the anomalous Green's function is proportional to the phase of the order parameter. In the present case of an atomically periodic phase modulation, this relation is not a priori obvious. However, our calculation shows that even here, the phase of the anomalous Green's function on a given sublattice remains proportional to the phase of the order parameter on that sublattice. Therefore, to remove this additional phase factor and isolate the pure conventional (homogeneous) and N\'eel triplet components, we divide the triplet anomalous Green's function by the (complex) order parameter on the corresponding sublattice. The resulting quantities are purely real and can thus be decomposed into conventional and N\'eel components.

For $h_{A,B,C}=D\{+8,-10.5,0\}$~meV, the magnitude of triplet correlations varies only weakly among the $A$, $B$, and $C$ sublattices, indicating that conventional triplet correlations dominate over N\'eel ones in this case. Figure~\ref{fig:S1}b shows the difference in triplet correlations between sublattices $A$ and $B$, defined as $\tilde F_N^t = F_A^t/(i\Delta_A) - F_B^t/(i\Delta_B)$, which represents the N\'eel triplet component. Its smallness stems from two factors: (i) the effective exchange field is ferrimagnetic rather than antiferromagnetic, and (ii) in our case $|h_{A,B,C}| \ll t_0$, corresponding to the regime of strong suppression of N\'eel triplet correlations \cite{Bobkov2023_impurities}.

For comparison, Fig.~\ref{fig:S2} shows the triplet correlation amplitudes for all three sublattices in the case of a purely antiferromagnetic exchange field $h_{A,B,C}=D\{8, -8, 0\}$~meV. Here, the conventional homogeneous triplet component is completely absent. Only the N\'eel triplet component survives, with a magnitude comparable to that in the previous case. This smallness of the N\'eel triplets is due to the same physical reason: the antiferromagnetic exchange field is small compared to the on-site energy.

Finally, Fig.~\ref{fig:S3} presents a schematic (not to scale) of the spatial distribution of the triplet correlation amplitude for both cases considered.

\end{widetext}

\bibliography{MnPS3NbSe2}

\end{document}